\documentclass{PoS}

\usepackage{graphicx}
\usepackage{amsmath}
\usepackage{amssymb}
\newcommand{\be}{\begin{equation}}
\newcommand{\ee}{\end{equation}}
\newcommand{\bea}{\begin{eqnarray}}
\newcommand{\eea}{\end{eqnarray}}
\newcommand{\bi}{\begin{itemize}}
\newcommand{\ei}{\end{itemize}}
\newcommand{\ben}{\begin{enumerate}}
\newcommand{\een}{\end{enumerate}}
\newcommand{\bt}{\begin{tabbing}}
\newcommand{\et}{\end{tabbing}}

\title{
   \begin{picture}(0,0)(0,0)%
   \end{picture}%
Critical end point of Nf=3 QCD at finite temperature and density
}

\ShortTitle{
Critical end point of Nf=3 QCD at finite temperature and density
}

\author{
   \speaker{Shinji Takeda}$^{a,b}$\thanks{E-mail: takeda@hep.s.kanazawa-u.ac.jp}, 
   Xiao-Yong Jin$^b$,
   Yoshinobu Kuramashi$^{b,c,d}$,
   Yoshifumi Nakamura$^b$,
   and
   Akira Ukawa$^b$
   \\
   \\
   \\
   \llap{$^a$}
Institute of Physics, Kanazawa University, Kanazawa 920-1192, Japan
   \\
   \llap{$^b$}
RIKEN Advanced Institute for Computational Science,
Kobe, Hyogo 650-0047, Japan
   \\ 
   \llap{$^c$}
Graduate School of Pure and Applied Sciences,
University of Tsukuba,
Tsukuba, Ibaraki 305-8571, Japan
   \\
   \llap{$^d$}
Center for Computational Sciences,
University of Tsukuba,
Tsukuba, Ibaraki 305-8577, Japan
}
   
\abstract{
We investigate the phase structure of 3-flavor QCD in the
presence of finite quark chemical potential $a\mu=0.1$ by using the Wilson-Clover fermion action.
Especially, we focus on locating the critical end point that characterizes the phase structure. 
We do this by the kurtosis intersection method for the quark condensate.
For Wilson-type fermions, the correspondence between bare parameters and
physical parameters is indirect. 
Hence we present
a strategy to transfer the bare parameter phase structure to the physical one.
}

\FullConference{32nd International Symposium on Lattice Field Theory LATTICE 2014\\
		 June 23 - 28, 2014\\
		 New York, USA}

\begin{document}

\section{Introduction}

At zero baryon number density,  in the two-dimensional plane spanned
by the light (up-down degenerated) quark mass $m_{ud}$ and strange quark mass $m_{s}$,
the first order phase transition around the massless point $m_{ud}=m_s=0$ 
becomes weaker as the quark masses increase, 
and eventually turns into a crossover at some  finite quark masses.
The boundary between the first order phase transition region and
the crossover region forms a line of second order phase transition,  and it is called
the critical end line.

Monte Carlo results on the location of the critical end line is rather confusing at present.  For the staggered fermion 
action, recent studies with improved action could place only an upper bound on the critical quark mass, which is very small in the range of $m/m_{ud}^{phys}\approx 0.1$~\cite{Endrodi:2007gc,Ding:2011du}.  
This is  in contrast to an earlier work with naive action~\cite{Kayaetal1999,Karsch:2001nf} which observed first order signals up to $m/m_{ud}^{phys}\approx 2-3$.   
Furthermore, our recent study with the Wilson-clover fermion action~\cite{Nakamura} could identify the critical end point, although the cut-off dependence of the location is rather large.

Another important issue with the QCD phase diagram is how the critical end line extends when switching on the chemical potential.
An interesting result was reported in \cite{deForcrand:2006pv}
which explored the imaginary chemical potential approach with the staggered fermion action. 
There it was observed that the critical surface has a negative curvature in the $\mu$ direction although the error is large.

Our purpose in this report is to investigate the critical end point at zero and finite density along the $N_f=3$ symmetric line.  In particular, we investigate the curvature of the critical end point by using the Wilson-clover fermion action, and study the signature of the curvature.

\section{Strategy for locating the critical end point}

\begin{figure}
\begin{center}
\begin{tabular}{cc}
\hspace{-5mm}
%\vspace{8mm}
\scalebox{0.4}{\includegraphics{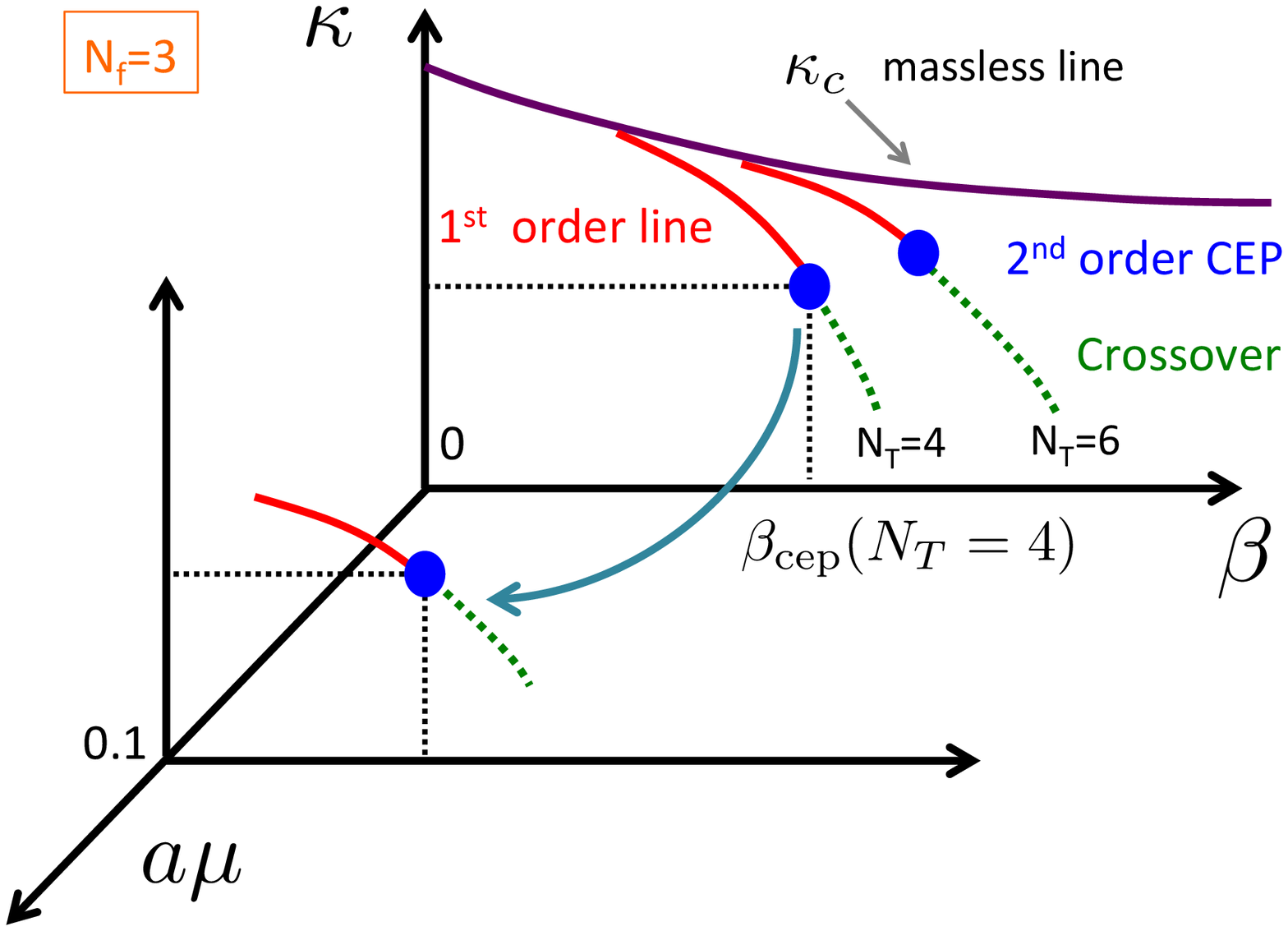}}
&
\hspace{10mm}
\scalebox{0.4}{\includegraphics{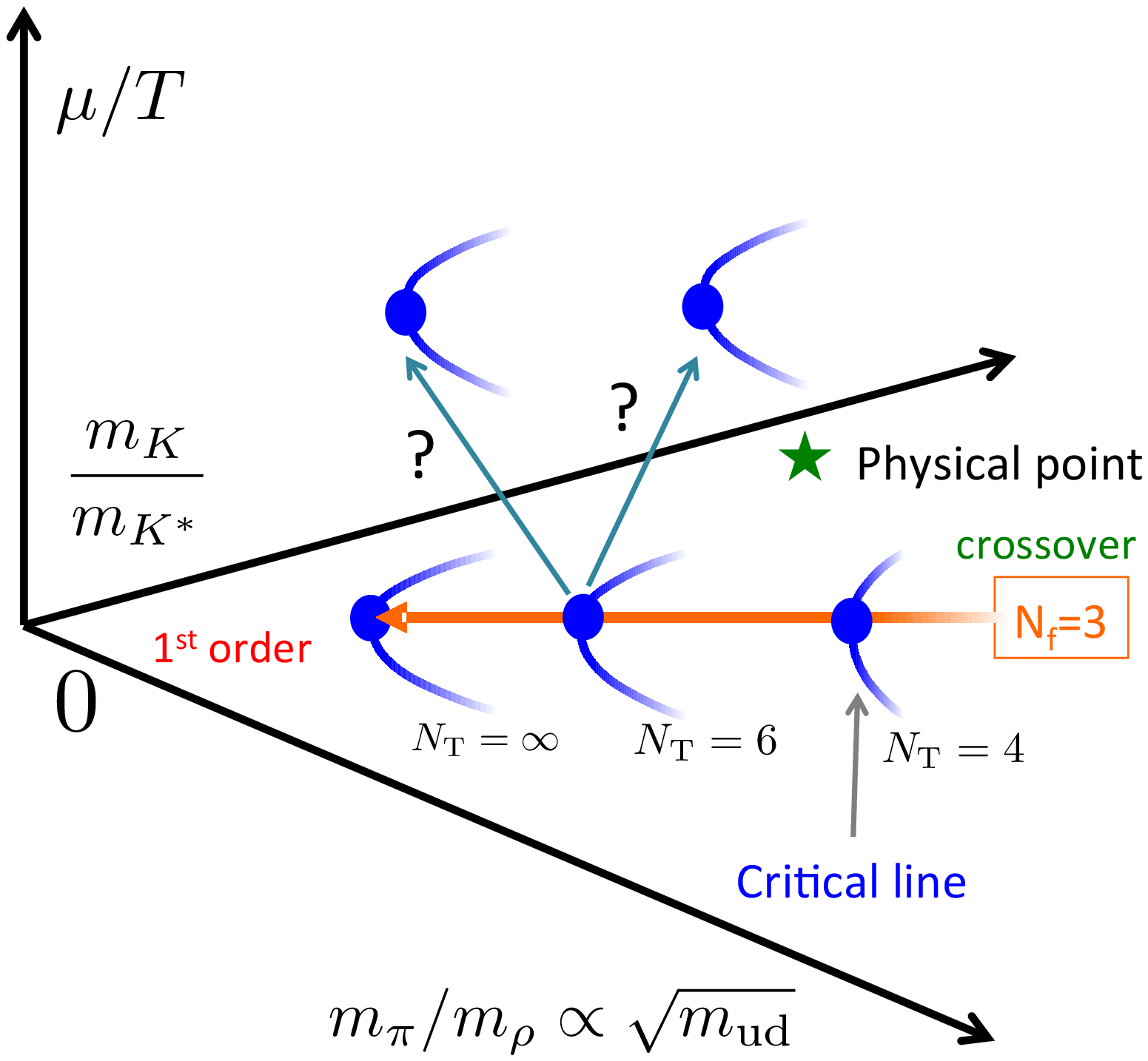}}
\\
\end{tabular}
\end{center}
\caption{
Strategy:
The left panel is the phase diagram for bare parameters spanned by $\beta$, $\kappa$ and $a\mu$
for $N_{\rm f}=3$.
The right panel is the same phase diagram but depicted for physical parameters spanned by $m_\pi/m_\rho$,
$m_K/m_{K^\ast}$ and $\mu/T$.
}
\label{fig:strategy}
\end{figure}

Let us discuss the strategy to survey the phase space and identify the critical end point for the 
Wilson-clover fermion action.  
If we consider the zero density case with  $N_{\rm f}=3$ degenerate quarks, 
we only have two bare parameters $\beta$ and $\kappa$
($a\mu=0$ plane in the left panel in Fig.~\ref{fig:strategy}).
For a given temporal lattice size, say $N_{\rm T}=4$, by using the peak position of susceptibility
or zero of skewness of quark condensate,
one can draw the line of  finite temperature transition (the red line in the left panel in Fig.~\ref{fig:strategy}).
The transition changes from being of first order to cross over at a second order critical end point (the blue point in the left panel in Fig.~\ref{fig:strategy}). 
We compute the kurtosis of quark condensate along the transition line for a set of spatial volumes. 
The  intersection point is identified as the critical end point \cite{Karsch:2001nf}.
In this way, we can determine the critical end point in the bare parameter space
$(\beta_{\rm cep},\kappa_{\rm cep})$
and this procedure can be repeated to another value of $N_{\rm T}$.

%%%%%%%%
%%%%%%%%%

In order to translate the critical end point in the bare parameter space to that 
in the physical parameter space, we measure
hadron mass ratios such as the pseudo scalar-vector meson  mass ratio,
$m_{\rm PS}/m_{\rm V}$ for the bare parameters $(\beta_{\rm cep},\kappa_{\rm cep})$
by a zero temperature simulation.
We do not use quark masses
to avoid the multiplicative renormalization issue.
In this way we obtain the phase structure in physical parameter space whose axes
are given by $m_\pi/m_\rho$, $m_K/m_{K^\ast}$ and $\mu/T$ as shown in the  right panel in Fig.~\ref{fig:strategy}. 
When discussing the $N_{\rm f}=3$ case with zero density,  we can move
along the diagonal line $m_\pi/m_\rho=m_K/m_{K^\ast}$ on the $\mu/T=0$ plane.
The critical end point in the physical space is denoted by the blue point in the figure.
By repeating the same calculation for increasingly larger values of $N_{\rm T}$,
we can take the continuum of limit of the critical end point in the physical parameter space,
\be
\lim_{N_{\rm T}\rightarrow\infty}
\frac{m_{\rm PS,cep}}{m_{\rm V,cep}}(\mu/T=0).
\ee

When switching on the chemical potential,
the basic procedure is the same;  one just has to repeat the same analysis on a  different plane with $\mu\neq0$. 
Our final goal is to see whether the critical surface in the right panel in Fig.~\ref{fig:strategy}
bends toward the lighter mass or heavier mass direction.

\section{Simulation details}
We employ the Wilson-clover fermion action with non-perturvatively tuned $c_{\rm sw}$~\cite{CPPACS2006}
and Iwasaki gauge action.
The number of flavor is three, $N_{\rm f}=3$, and
the masses and chemical potentials for quarks are all degenerate.
The phase reweighting method is adopted to handle the complex phase according to 
\be
\langle
{\cal O}
\rangle
=
\frac{
\langle
{\cal O} e^{iN_{\rm f}\theta}
\rangle_{||}
}
{
\langle
e^{iN_{\rm f}\theta}
\rangle_{||}
},
\ee
where $
\langle
...
\rangle_{||}
$ is the average with phase quenched fermion determinant
\be
{\cal Z}_{||}
=
\int [dU]
e^{-S_{\rm G}}
|\det D_{\rm w}|^{N_{\rm f}},
\ee
and the phase factor is given by
\be
e^{i\theta}
=
\frac{\det D_{\rm w}}{|\det D_{\rm w}|}.
\ee
Configurations are generated by RHMC
with the phase quenched quark determinant.  The phase factor is computed exactly 
using the analytical reduction technique
\cite{Danzer:2008xs,Takeda:2011vd}. 
The dense matrix obtained by the reduction is computed
on GPGPU with LAPACK routines.

The temporal lattice size and the quark chemical potential
are fixed to $N_T=6$ and $\mu a=0.1$,  respectively, and thus $\mu/T=0.6$. 
To perform  finite size scaling analyses, the spatial volume is changed from $8^3$ to $12^3$.
In order to search for the transition point,
we select four $\beta$ points, namely, $\beta=1.70$, $1.73$, $1.75$ and $1.77$,
and  for each $\beta$, we vary $\kappa$ to locate the transition point.

We measure the quark condensate
and its higher moments up to fourth order at every 10th  trajectory.
In the computation of trace of higher powers of inverse of the lattice Dirac operator,
we adopt the noise method with 20 Gaussian noises for all parameter sets.
We have checked that number is sufficient to control the noise error.

\section{Simulation results}

\subsection{Phase reweighing factor}

\begin{figure}[t]
\begin{center}
\begin{tabular}{c}
\scalebox{1.4}{\includegraphics{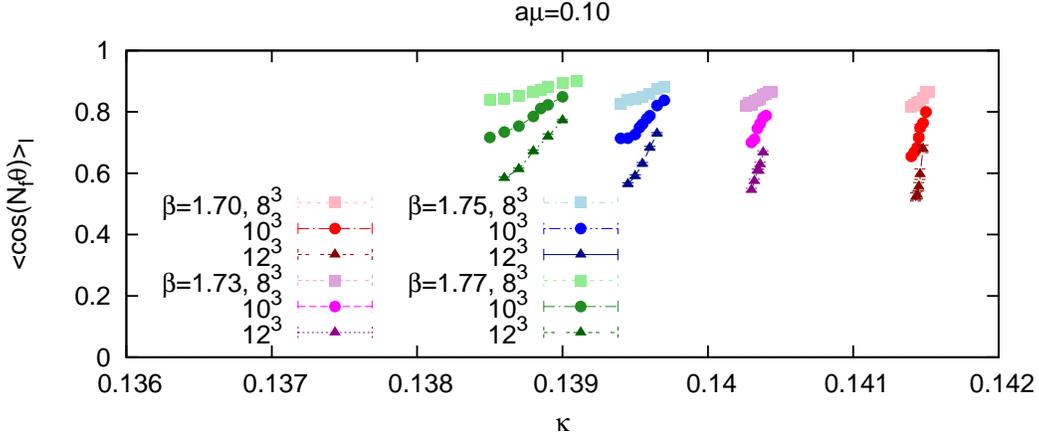}}
\end{tabular}
\end{center}
\caption{Average of phase reweighting factor as a function of $\kappa$.
$\mu/T=a\mu\times N_{\rm T}=0.1\times6=0.6$.
The sign problem is mild in this region.
}
\label{fig:phasereweighting}
\end{figure}

Figure~\ref{fig:phasereweighting} shows the average value of the phase-reweighting factor as a function of $\kappa$.
For small $\kappa$ and large volumes, the value tends to be close to zero, 
signaling that the sign problem is becoming serious.
Nevertheless, it stays away from zero beyond statistical error, guaranteeing the viability of 
reweighing for our range of lattice volumes.

%%%%%%%%
%%%%%%%%%

\subsection{Kurtosis intersection}

\begin{figure}[t]
\begin{center}
\begin{tabular}{c}
\scalebox{0.9}{\includegraphics{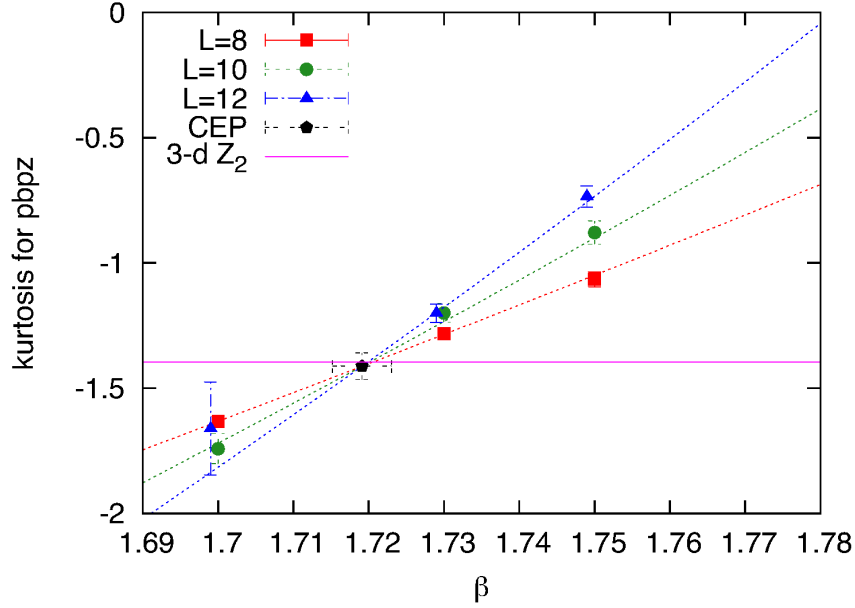}}
\end{tabular}
\end{center}
\caption{
Kurtosis intersection for quark condensate.
Lines show fitting function in the text.
Black pentagon is the estimated critical end point, $\beta_{\rm cep}=1.7191(39)$ where 
critical value of kurtosis is $K_{\rm cep}=-1.413(53)$.
Horizontal magenta line shows the critical value of kurtosis
for 3-dimensional Ising model, $K_{\rm cep}=-1.396$.
}
\label{fig:kurtosis}
\end{figure}

The kurtosis for quark condensate
along the transition line  is plotted in Fig.~\ref{fig:kurtosis} as a function of $\beta$.
For a first order phase transition, the infinite volume value of kurtosis
is $K=-2$ while for a crossover it is $K=0$, 
which can be used to diagnose the strength of phase transitions.
The figure then shows that
a strong first order phase transition at lower $\beta$
becomes weaker for higher $\beta$.

At the critical end point, the kurtosis is expected to take the same value irrespective of the spatial volume 
between -2 and 0.  The value depends on the universality class of the second order phase transition at the critical end point.  Using the fitting form
\cite{deForcrand:2006pv} inspired by finite size scaling,
\be
K=K_{\rm cep}
+CL^{1/\nu}(\beta-\beta_{\rm cep}),
\label{eqn:Kintersectionformula}
\ee
where $K_{\rm cep}$, $C$, $\nu$ and $\beta_{\rm cep}$ are fitting parameters,
we determine the critical end point $\beta_{\rm cep}=1.7191(39)$,
the exponent $\nu=0.66(13)$ and the value of kurtosis at the critical end point $K_{\rm cep}=-1.413(53)$.
The values of the last  two parameters  are consistent with those of the 
3-dimensional Ising universality class, $\nu=0.63$ and $K_{\rm cep}=-1.396$.  On the other hand, 
the  universality class of 3-dimensional O(2) symmetry with $K_{\rm cep}=-1.758$
and that of  3-dimensional O(4) with $K_{\rm cep}=-1.908$ are rejected.

%%%%%%%%
%%%%%%%%%

\subsection{Susceptibility}
\begin{figure}[h]
\begin{center}
\begin{tabular}{cc}
\hspace{-10mm}
\scalebox{0.9}{\includegraphics{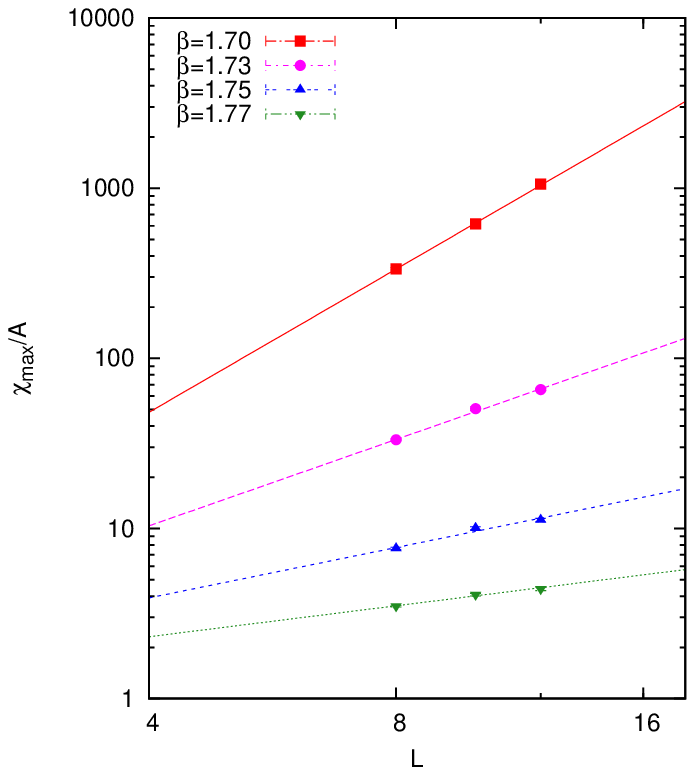}}
&
\hspace{-5mm}
\scalebox{1.2}{\includegraphics{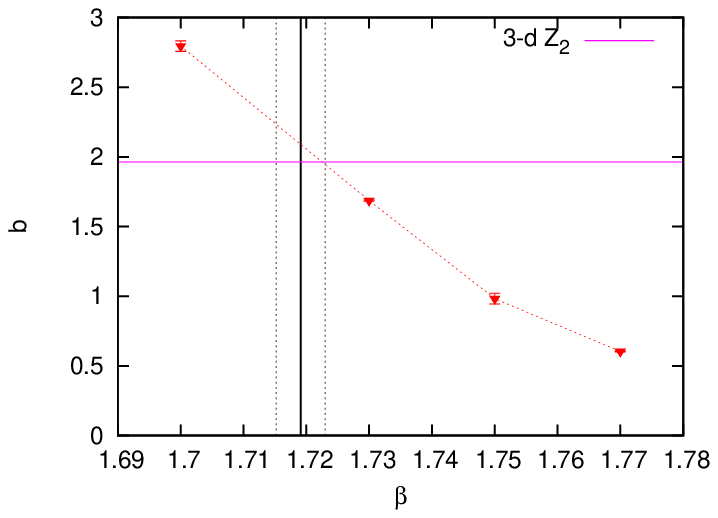}}
\end{tabular}
\end{center}
\caption{Left: Volume scaling of the susceptibility peak height for quark condensate
with fitting form $\chi_{\rm max}=AL^b$.
Note that both axes are log scale.
Right: Exponent $b$ along the transition line.
The vertical black line shows the critical end point
$\beta_{\rm cep}=1.7191(39)$.
The horizontal magenta line shows
the value of $\gamma/\nu=1.9630$ for 3-dimensional Ising model.
}
\label{fig:susceptibility}
\end{figure}

In order to confirm the universality class,
we check  another critical exponent $\gamma/\nu$. 
The left panel of Fig.~\ref{fig:susceptibility}
shows the volume scaling of the susceptibility peak height of quark condensate
together with fitting lines of  form
\be
\chi_{\rm max}=AL^b,
\ee
where $A$ and $b$ are fit parameters.
For lower $\beta$ the volume sensitivity is strong while
it becomes weaker for higher $\beta$. 
This signifies that the phase transition weakens by creasing $\beta$.

The  exponent $b$ obtained by the fit  is plotted along the transition line as a function of $\beta$
in the right panel of Fig.~\ref{fig:susceptibility}.
The value $b=1$ corresponds to a first order phase transition, while $b\approx0$ shows a crossover.
At the critical point where the second order phase transition takes place, 
the exponent is expected to be given by $b=\gamma/\nu$.
We observe that the exponent $b$ at the critical end point estimated by the kurtosis intersection
is consistent with the value for the 3-dimensional Ising model $\gamma/\nu=1.9630$.

%%%%%%%%
%%%%%%%%%

\subsection{Curvature of critical line}

\begin{figure}[h]
\begin{center}
\begin{tabular}{cc}
\scalebox{0.7}{\includegraphics{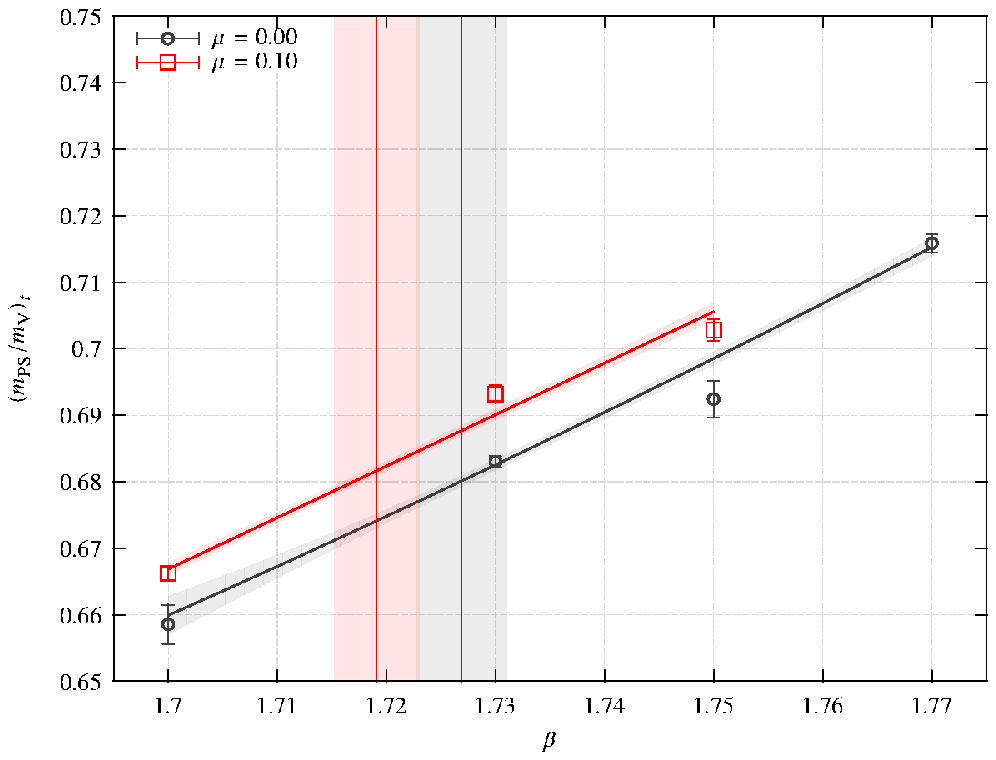}}
&
\scalebox{0.6}{\includegraphics{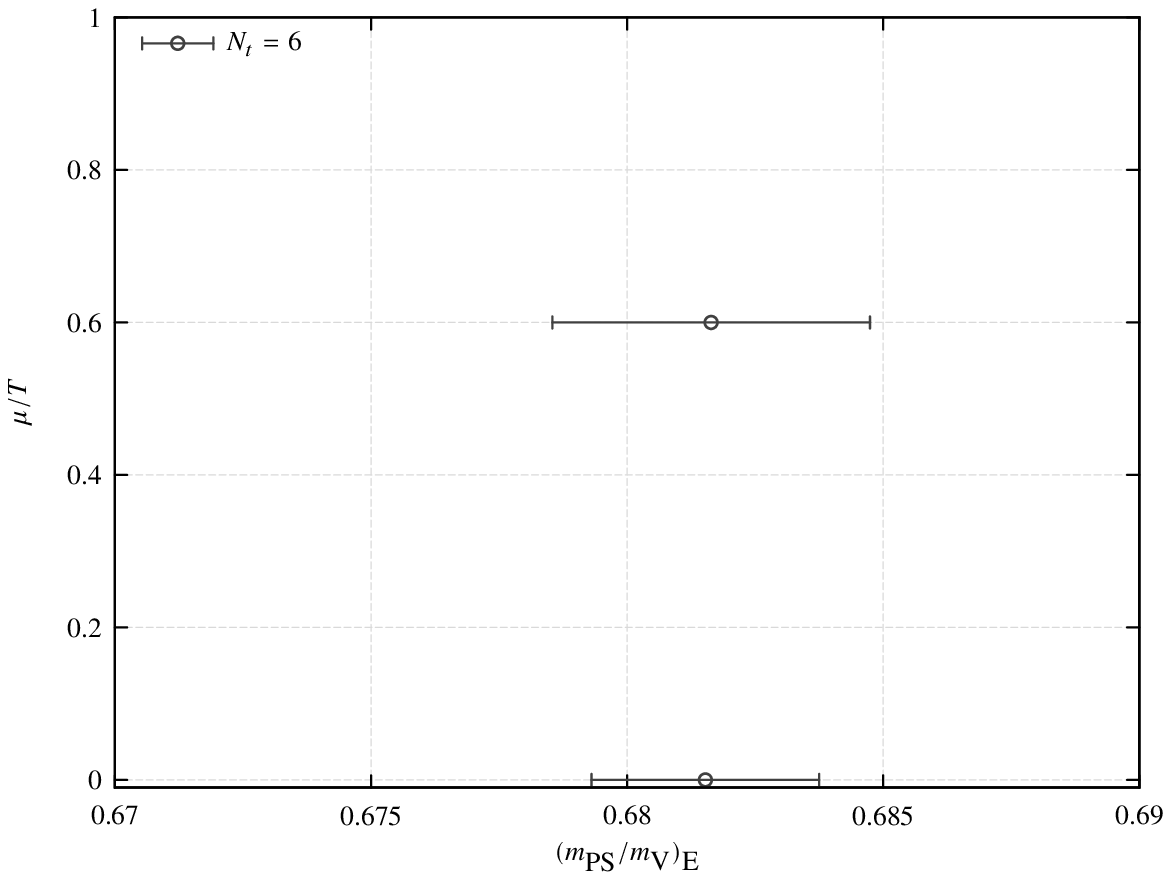}}
\end{tabular}
\end{center}
\caption{
Left: $m_{\rm PS}/m_{\rm V}$ along the transition line.
The black (red) points are for zero density ($a\mu=0.1$)
The black (red) vertical line shows
the location of critical end point for zero density ($a\mu=0.1$).
Right:
$\mu$-dependence of critical end point in the physical space (the hadron mass ratio).
The curvature of the critical line is too small to see at this scale.
}
\label{fig:CEP}
\end{figure}

In the previous subsection, we have determined the critical end point\footnote{
The corresponding $\kappa_{\rm cep}$ is also determined by
the transition line in $\kappa$-$\beta$ plane.
}
$\beta_{\rm cep}$.  
The last step is to translate the critical end point to the physical space.
The left panel of Fig.~\ref{fig:CEP} shows
the hadron mass ratio,
$m_{\rm PS}/m_{\rm V}$ along the transition line.
Since the transition line depends on the chemical potential,
two lines are shown in the panel for  zero density and for $a\mu=0.1$.  
The location of the critical end point $\beta_{\rm cep}$ is also shown as vertical lines; 
for zero density the estimate is taken from \cite{Nakamura}.
From this graph, we can read off the value of hadron mass ratio
at the critical end point.
In the right panel of Fig.~\ref{fig:CEP},
the $\mu$-dependence of
the hadron mass ratio at the critical end point is shown.
Even though the hadron mass ratio is determined very precisely with errors of less than 1\%,
no significant $\mu$-dependence is observed.
This results suggests that the curvature of critical surface is quite small.

%%%%%%%%
%%%%%%%%%

In future we shall try to reduce the errors in the final plot for both zero
and finite density using the multi-ensemble reweighting method.
Furthermore we shall explore the region beyond $a\mu=0.1$ by using the multi-parameter ($\mu$ and $\kappa$) reweighting method.
By reducing the errors and reaching the higher density region, we hope to answer the question raised in the introduction of the sign of the curvature of critical surface.

%// Acknowledgments -----------------------------------------------------------

\vspace{3mm}
This work is supported in part by the Grants-in-Aid for
Scientific Research from the Ministry of Education, 
Culture, Sports, Science and Technology 
(Nos.
23740177, %takeda wakate B
26800130). %takeda wakate B H26-
We acknowledge the CCS (Center for Computational Sciences in University of Tsukuba) simulation program
for allocating run time of HA-PACS.


\begin{thebibliography}{99}


%\cite{Endrodi:2007gc}
\bibitem{Endrodi:2007gc} 
  G.~Endrodi, Z.~Fodor, S.~D.~Katz and K.~K.~Szabo,
  %``The Nature of the finite temperature QCD transition as a function of the quark masses,''
  PoS LAT {\bf 2007}, 182 (2007)
  [arXiv:0710.0998 [hep-lat]].
  %%CITATION = ARXIV:0710.0998;%%
  %34 citations counted in INSPIRE as of 27 Oct 2014
  
%\cite{Ding:2011du}
\bibitem{Ding:2011du} 
  H.-T.~Ding, A.~Bazavov, P.~Hegde, F.~Karsch, S.~Mukherjee and P.~Petreczky,
  %``Exploring phase diagram of $N_f=3$ QCD at $\mu=0$ with HISQ fermions,''
  PoS LATTICE {\bf 2011}, 191 (2011)
  [arXiv:1111.0185 [hep-lat]].
  %%CITATION = ARXIV:1111.0185;%%
  %7 citations counted in INSPIRE as of 27 Oct 2014
  
%\cite{Kaya:1999}
\bibitem{Kayaetal1999}
JLQCD Collaboration (S. Aoki (Tsukuba U.) et al.), 
%{\it Phase structure of lattice QCD at finite temperature for 2+1 flavors of Kogut-Susskind quarks},  
Nucl. Phys. Proc. Suppl. {\bf 73},  459 (1999).

%\cite{Karsch:2001nf}
\bibitem{Karsch:2001nf} 
  F.~Karsch, E.~Laermann and C.~Schmidt,
  %``The Chiral critical point in three-flavor QCD,''
  Phys.\ Lett.\ B {\bf 520}, 41 (2001)
  [hep-lat/0107020].
  %%CITATION = HEP-LAT/0107020;%%
  %96 citations counted in INSPIRE as of 14 Nov 2013
    
\bibitem{Nakamura}
X-Y. Jin, Y. Kuramashi, Y. Nakamura, S. Takeda and A. Ukawa
PoS(Lattice 2014)


%\cite{deForcrand:2006pv}
\bibitem{deForcrand:2006pv} 
  P.~de Forcrand and O.~Philipsen,
  %``The Chiral critical line of N(f) = 2+1 QCD at zero and non-zero baryon density,''
  JHEP {\bf 0701}, 077 (2007)
  [hep-lat/0607017].
  %%CITATION = HEP-LAT/0607017;%%
  %199 citations counted in INSPIRE as of 27 Oct 2014
  
  

%\cite{CPPACS2006}
\bibitem{CPPACS2006}
CP-PACS and JLQCD Collaborations (S. Aoki {\it et al.}), Phys. Rev. D{\bf 73}, 034501 (2006)
[hep-lat/0508031].

%\cite{Danzer:2008xs}
\bibitem{Danzer:2008xs} 
  J.~Danzer and C.~Gattringer,
  %``Winding expansion techniques for lattice QCD with chemical potential,''
  Phys.\ Rev.\ D {\bf 78}, 114506 (2008)
  [arXiv:0809.2736 [hep-lat]].
  %%CITATION = ARXIV:0809.2736;%%

%\cite{Takeda:2011vd}
\bibitem{Takeda:2011vd} 
  S.~Takeda, Y.~Kuramashi and A.~Ukawa,
  %``On the phase of quark determinant in lattice QCD with finite chemical potential,''
  Phys.\ Rev.\ D {\bf 85}, 096008 (2012)
  [arXiv:1111.6363 [hep-lat]].
  %%CITATION = ARXIV:1111.6363;%%
  %3 citations counted in INSPIRE as of 28 Oct 2014
  

  

\end{thebibliography}
\end{document}